# A Machine Learning Application for Raising WASH Awareness in the Times of COVID-19 Pandemic.

## Original Paper

Rohan Pandey[1 #]; Vaibhav Gautam[1 $]; Ridam Pal[2 $]; Harsh Bandhey[2 $]; Lovedeep Singh Dhingra[2,3 $]; Himanshu Sharma[4 &]; Chirag Jain[2 &]; Kanav Bhagat[2 &]; Arushi[3 &]; Lajjaben Patel[3 &]; Mudit Agarwal[3 &]; Samprati Agrawal[3 &]; Rishabh Jalan[2 &]; Akshat Wadhwa[2 &]; Ayush Garg[2 &]; Vihaan Misra[5 &]; Yashwin Agrawal[2 &]; Bhavika Rana[2]; Ponnurangam Kumaraguru[2]; Tavpritesh Sethi[2]

[1]Shiv Nadar University, Uttar Pradesh, Noida, India
[2]Indraprastha Institute of Information Technology, Delhi, India
[3]All India Institute of Medical Sciences, New Delhi, India, India
[4]GL Bajaj Institute Of Tech and Management, Uttar Pradesh, Greater Noida, India
[5]Netaji Subhas University of Technology, Dwarka, New Delhi, India

[#] contributed equally, [$] contributed equally, [&] contributed equally

Corresponding Author:
Tavpritesh Sethi
Department of Computational Biology
Indraprastha Institute of Information Technology, Delhi
Okhla Industrial Estate, Phase III, New Delhi - 110020
Delhi
India
Email: tavpritesthsethi@iiitd.ac.in

## Abstract

**Background:** The COVID-19 pandemic has uncovered the potential of digital misinformation in shaping the health of nations. The deluge of unverified information that spreads faster than the epidemic itself is an unprecedented phenomenon that has put millions of lives in danger. Mitigating this 'Infodemic' requires strong health messaging systems that are engaging, vernacular, scalable, effective and continuously learn the new patterns of misinformation.
**Objective:** We created WashKaro, a multi-pronged intervention for mitigating misinformation through conversational AI, machine translation and natural language processing. WashKaro provides the right information matched against WHO guidelines through AI, and delivers it in the right format in local languages.
**Methods:** We theorize (i) an NLP based AI engine that could continuously incorporate user feedback to improve relevance of information, (ii) bite sized audio in the local language to improve penetrance in a country with skewed gender literacy ratios, and (iii) conversational but interactive AI engagement with users towards an increased health awareness in the community.
**Results:** A total of 5026 people who downloaded the app during the study window, among those 1545 were active users. Our study shows that 3.4 times more females engaged with the App in Hindi as compared to males, the relevance of AI-filtered news content doubled within 45 days of continuous machine learning, and the prudence of integrated AI chatbot "Satya" increased thus proving the usefulness of an mHealth platform to mitigate health misinformation.
**Conclusion:** We conclude that a multi-pronged machine learning application delivering vernacular bite-sized audios and conversational AI is an effective approach to mitigate health misinformation.

**Trial Registration:** Not Applicable

## Introduction

Healthcare misinformation is a growing menace in digital societies. This is clearly highlighted by the COVID-19 pandemic that has affected over 3.8 million people worldwide causing a widespread loss in all aspects of daily life [1]. Digital consumption has increased manifolds, creating both an opportunity and a danger in terms of information dissemination. Infodemic has been defined as an overabundance of information, some accurate and some not, that makes it hard for people to find trustworthy sources and reliable guidance when they need it [2]. The spread of COVID-19 infodemic was much faster than the pandemic itself and still continues to pose a threat to public health [3]. Further, mitigation of misinformation is also important for raising correct awareness for primary prevention of most communicable and non-communicable diseases. However, the role of machine learning, artificial intelligence and digital technologies in mitigating health misinformation has been under-explored. Mobile health (mHealth), coupled with verified health information, can serve as an information dispensing tool to tackle the spread of misinformation. Clear and effective communication of preventive measures and updated information is essential. To achieve this goal, designing a trustworthy app that helps navigate the information deluge is important. Therefore, recognizing the potential of mHealth platforms, we developed WashKaro, a multi-pronged AI approach for Infodemic Management. WashKaro was driven by the imminent need to raise Water, Sanitation, and Hygiene (WASH) awareness and combines English (WASH) with vernacular (Karo, meaning "Do" in Hindi) for mitigating the spread of COVID-19. WashKaro combines Natural Language Processing (NLP) to match information with WHO guidelines(OnAIr), and conversational AI (Satya, meaning "Truth" in Hindi) to reach out to the community as audio-visual content in local languages. In order to keep the information relevant, WashKaro provides daily news matched with WHO guidelines, WHO directive-based Symptom Self-Assessment tool, and human-vetted information delivering these in Hindi, the most widely understood local language across India. Since India is one of the largest and fastest-growing markets for digital consumers, with 560 million

Internet subscribers in 2018 [4], and about 60% using mHealth technologies [5], this offered a unique opportunity to test WashKaro.

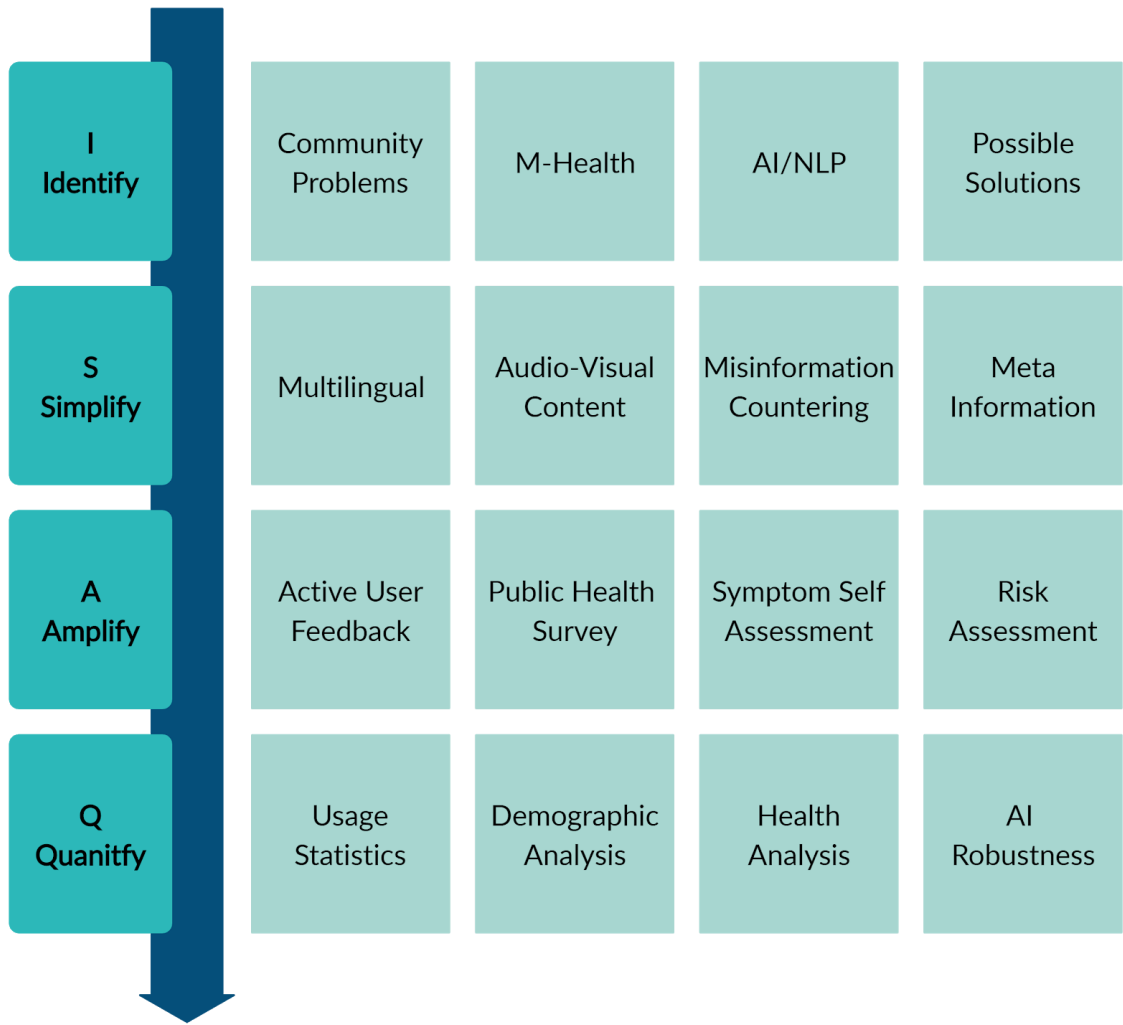

Figure 1: Our proposed workflow. The methodology used for this study is centered around the WHO EPI-WIN strategy which covers four strategic areas of work to respond to the infodemic. The first area of coverage is focused on identifying the the problem at hand given the current evidence and information in order to strategically promote and form public policies. We have identified the context-specific community problems and the potential of mHealth, AI, and NLP in order to acknowledge possible solutions. This is followed by the simplification of the enormous amount of information currently available across multiple sources in order to disseminate accurate information in a simplified manner. In order to achieve this objective, multilingual support is provided in the form of audio visual-based content. The spread of misinformation is tackled by providing information such as Mythbusters and government updates along with meta-information in the form of geographic coordinates of essential facilities. We also provide periodic hand washing reminders. In case of such an unforeseen event, it is important to amplify the intervention by means of establishing two-way communication with the intended audience to tailor the advice and messages. This has been catered to by engaging users in active feedback based involvement, participating in enhancing the AI proposed model along with any generic feedback in an audio format. A public health survey and symptom self-assessment are crucial components in amplifying our study. In order to devise constantly evolving strategies it is essential to validate the methodology and quantify the infodemic. WashKaro application statistics, demographic analysis of public health surveys, health analysis of at-risk population using symptom self-assessment, and user agreement on AI-based intervention is critical to quantify and evaluate.

## Methods

WashKaro was developed as a holistic mHealth solution that could serve as a one-stop AI-powered infodemic management suite during the current COVID 19 pandemic. The underlying strategy utilized was Identify-Simplify-Amplify-Quantify, as deployed by the Information Network for Epidemics (EPI-WIN) established by the WHO [2]. Gathering raw data from credible sources such as the WHO, and consumer centric daily news articles, we used NLP approaches along with machine learning to identify authentic and pertinent information. The information thus extracted was simplified and presented as audio-visual content in Hindi (the most widely understood local language across India), English, and various other vernaculars. By garnering feedback on the relevance of the WHO information provided along with the news pieces, the advice to the individual was tailored according to their personal needs, thus amplifying the reach of appropriate messages. We also offered a WHO directive-based Symptom Self-Assessment tool, as well as numerous categories of human-vetted information in the form of Infographics, MythBusters, geographic information, etc. Forming real-time, on-the-ground, multidisciplinary research partnerships is essential to mitigate the infodemic. Our entire methodology and infodemic suite is, therefore, open-source and available for the entire scientific community to build upon it.

### NLP in healthcare

In the current situation, timely delivery of tenable content to the masses is exceptionally crucial to counter the spread of misinformation. WashKaro targeted this requirement using Natural Language Processing techniques to dispense information sourced through highly trusted outlets of WHO such as EPI-WIN, which perhaps may not reach the appropriate audience, or be too complicated for them [6].

The NLP pipeline involved two datasets, namely the WHO guidelines and the news articles. After preliminary preprocessing of both the datasets, extractive ML summarization techniques were used to abbreviate the text. Embeddings were generated for the summarized guidelines and news articles and matched against each other. Subsequently, the users were provided with the pair of WHO guidelines and its matching news article with the highest similarity score. This pipeline served to complement the user's daily news consumption that suits their palette with an appropriate WHO guideline related to COVID-19 and WASH (Water Sanitation and Hygiene), thus augmenting healthcare awareness. In order to enhance engagement and provide increasingly relevant content, user feedback was sought at the end of each matching- the users marked each pair of WHO guideline and news article provided to them as either relevant or irrelevant. This active user feedback aided the machine learning backend in improving with each review by determining the type of news articles the user found relevant to a particular guideline. Further, any new article provided to the user took into account the previous learning, which enabled deliverance of more relevant information with each feedback cycle.

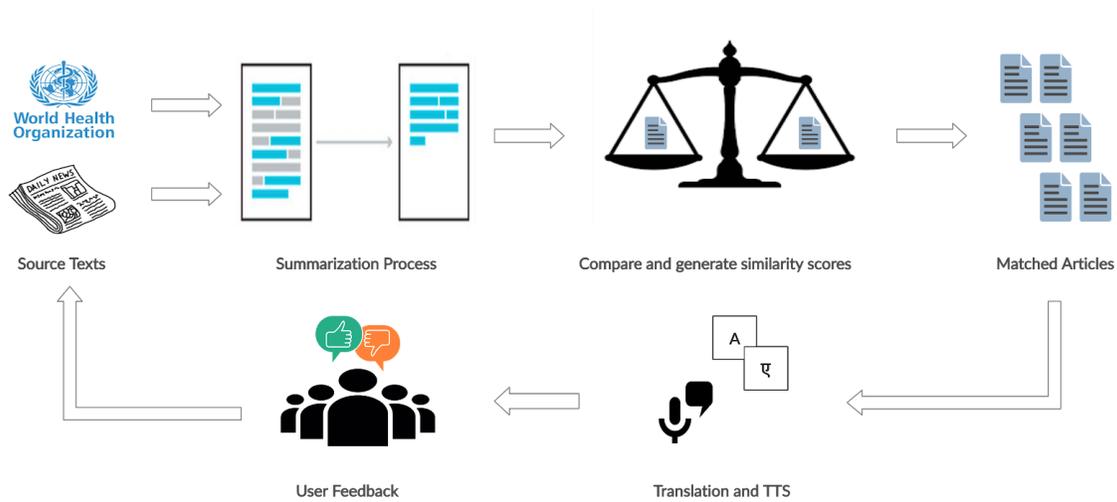

Figure 2: NLP Pipeline. The pipeline takes in news articles and WHO reports, and constructs two-level sentence similarity between titles and the full-text to construct a similarity score. Finally, the relevant texts are subject to translation and text to speech conversion for consumption in the local language (Hindi).

## Simplification

We made a deliberate attempt to convey context-specific and consumable information in a medium of the user's choice. Infographics based on WHO recommendations were used for effective presentation of preventive measures. Byte sized information packets were delivered in multiple local languages to ensure accessibility to various marginalized groups. Text-to-speech engines helped convert the information to an audio-visual format, thus reaching out to the less educated population. Mythbusters and government advisories, critical in countering misinformation and uncertainties surrounding the official guidelines, were credibly sourced and regularly updated. Information on containment zones, hospitals, and hunger relief centers was provided in a geographical context, with directions imparted through Google Maps, a popular user-friendly interface. Regular notifications, worded positively to encourage participation, reminding the user to wash their hands and use masks in public places, were displayed.

### Symptom Self Assessment

Low accessibility of the healthcare system, given the lockdown and social distancing measures in place, and a skewed ratio between the population who wishes to get tested and medical professionals who can verify this need, call for an effective alternative to screen patients [7]. Thus, we devised a self-assessment tool for the symptoms of COVID-19, thereby enabling quicker identification of suspect cases who can then be guided to the Government helpline numbers and informed about proper self-quarantine protocols, nearby hospitals admitting COVID suspects, and testing centers. We defined the Suspect Case using the WHO Interim Guidance on Global surveillance for COVID-19 caused by human infection with the COVID-19 virus, and classified them further as Suspect case (A), (B) or (C) [8]. The 7-point questionnaire was designed using the case definitions from the WHO Interim Guidance verbatim. Based on the application of the WHO criteria on the answers to the 7 questions, the user was notified about whether or not they were suspected of having COVID-19 (Fig. 3).

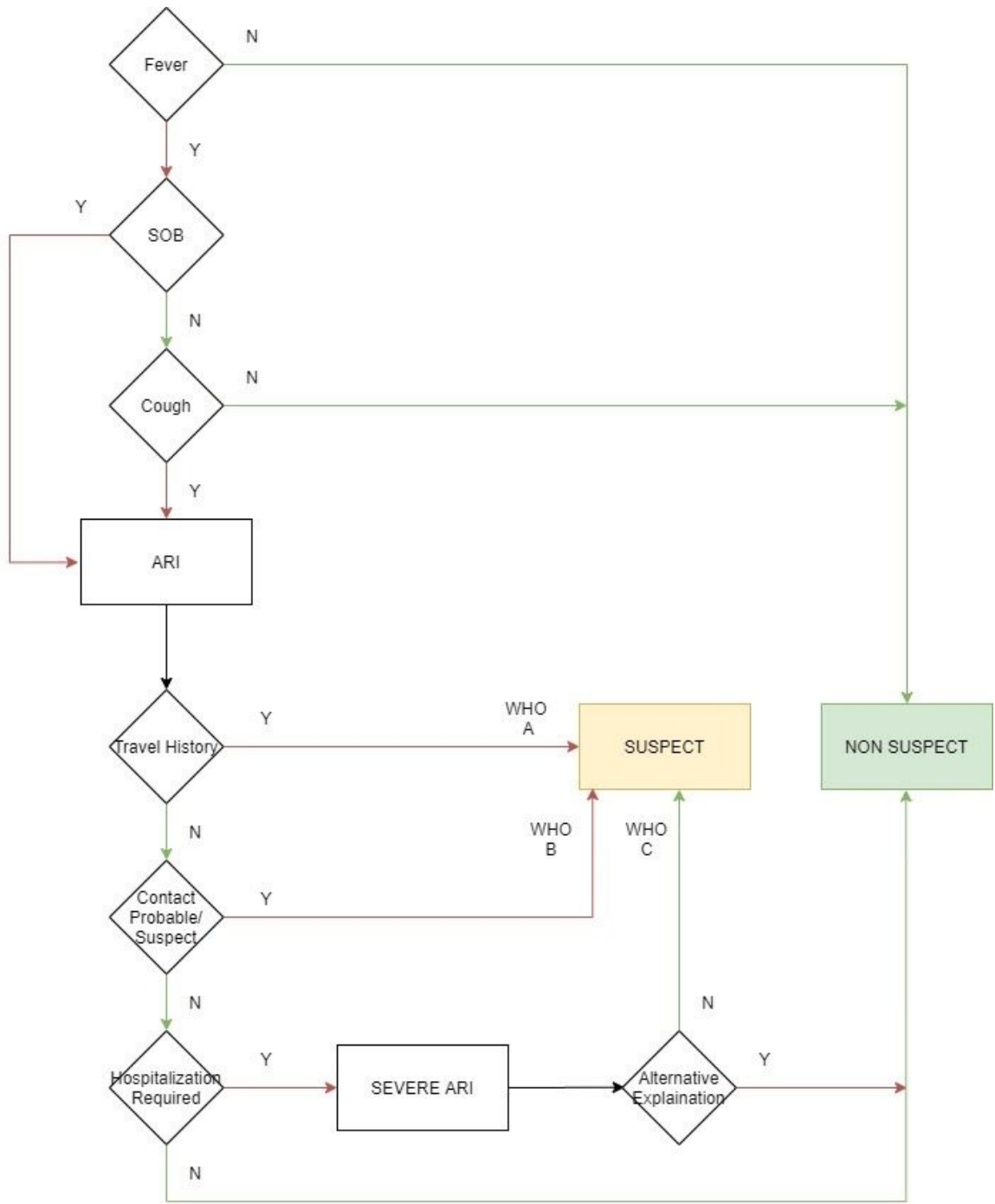

Figure 3. Self Assessment Tool Flowchart. Based on the WHO Interim Guidance, a questionnaire and flowchart were developed to classify the responders as `Suspects' or `Non-suspects'.

### Chatbot
Correct and officially verified information regarding the disease should be there for everyone's disposal. To serve this purpose, we have made a chatbot, which has verified information from WHO, CDC, and additional government-approved sources. Existing Solution consists of an option Driven System[9] where a user needs to select through various lists of options to find answers to the Query. Thus, we devised a chatbot system designed to answer user queries using natural language. The current system consists of an Lstm model fine-tuned on a MedSquad dataset[10]. The Dataset was encoded using swivel embeddings generated on the Covid-19 open research dataset[11]. All the data including the training set is incubated from Credible and Government controlled sources. The user Query is also passed through spelling correction using symmetric delete spelling correction algorithm along with artificial increasing frequency of words related to the disease, symptoms, etc.[12] so as to increase the accuracy and effectiveness of the system.

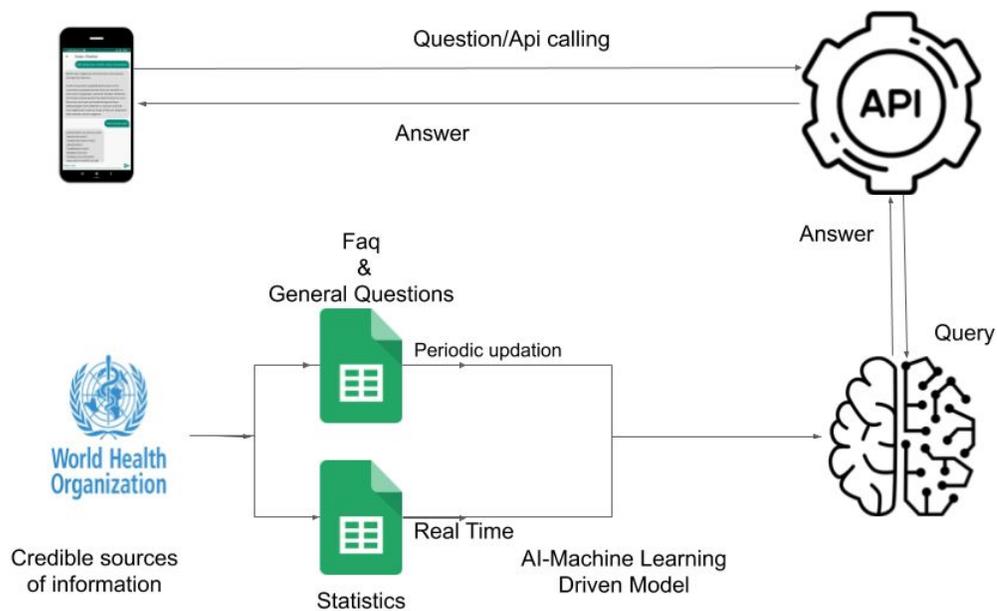

Figure 4. Request-Response cycle in the chatbot. This is a schematic diagram depicting how the answer is displayed whenever a query is asked to the chatbot by a user.

### Active User Feedback

Anonymized data was collected through self-assessment usage analytics, Play Store managed user statistics, and easy to comprehend survey forms. Our suite deployed an anonymized public health survey that asked basic healthcare-related questions in order to understand the demographics and to monitor the situation on a periodic basis. To ensure user convenience, establish a two-way dialogue and prevent specific suggestions from being marginalized, audio-based feedback was taken from the user. Illustrations were used whenever possible to make the user aware of the data being collected, hence protecting their right to information and privacy.

## Results

### Information Enrichment Over Time: The Number of `Relevant' Votes Increases

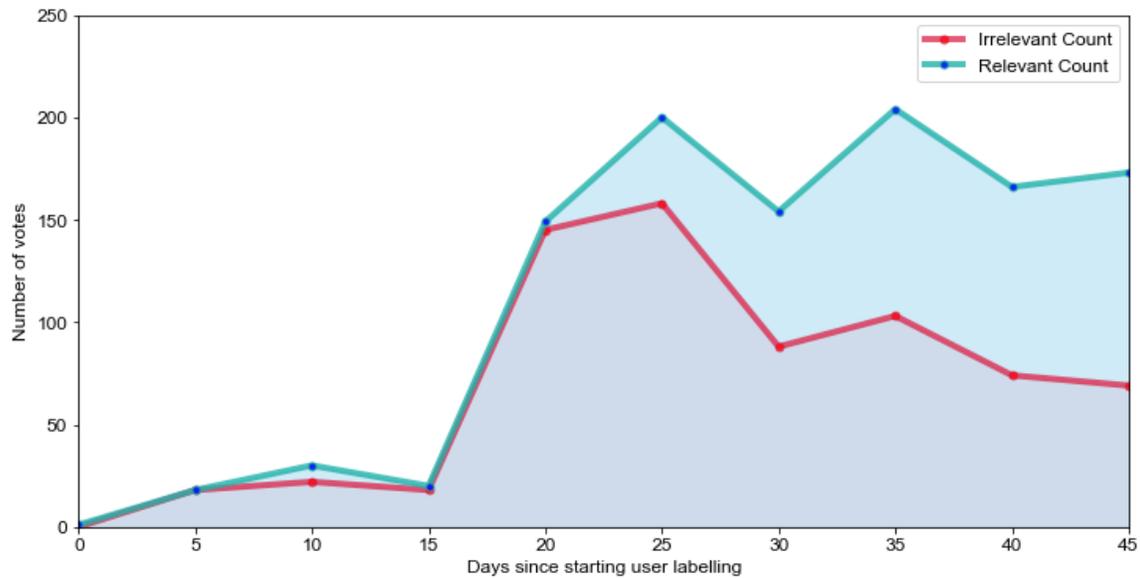

Figure 5. Analysis of NLP pipeline. Change in the number of relevant and irrelevant votes over time, with increasing feedback the relevance of the content increases.

With time and increased user feedback, the relevance of the matching of news articles with WHO reports increases as seen by the increase in the number of relevant votes, owing to the constantly evolving machine learning model. The number of irrelevant votes also decreases, validating our proposed methodology and providing increasingly relevant content from trusted sources to the user over time in the language of their preferred choice. At the beginning of the AI-based learning system on 15 March 2020, the number of `relevant' votes and `irrelevant' votes were both 18. On 25 March 2020, with an increase in user interaction and AI learning, the number of relevant votes was 173 and irrelevant votes were 69. The ratio of relevant votes to irrelevant votes increased from 1.0 to 2.5 over a period of one month.

### Demographics- Females engaged more in Hindi

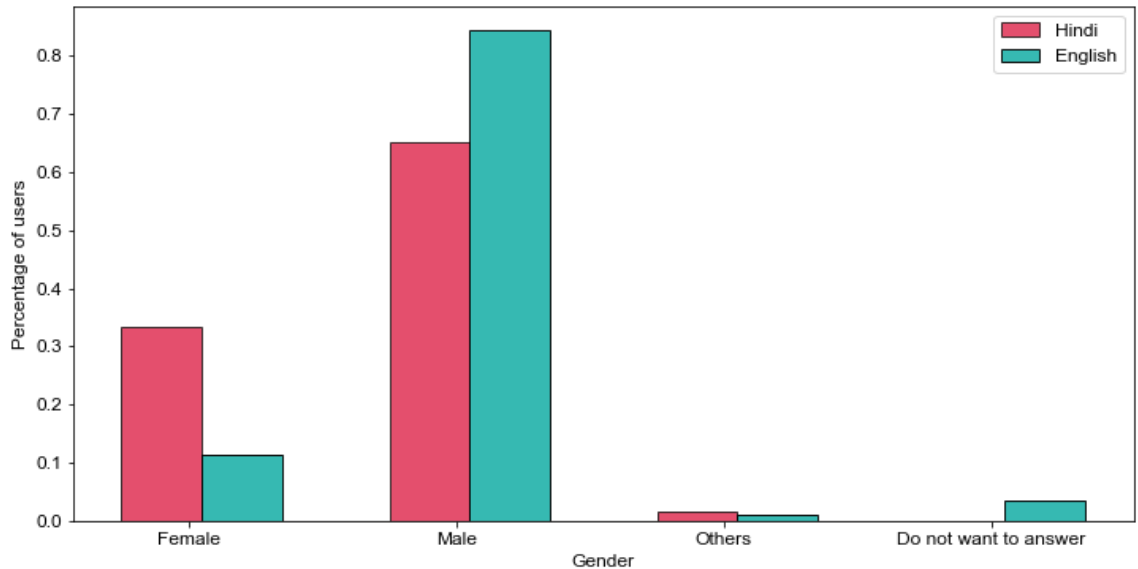

Figure 6. Analysis of Public Health Survey. Distribution graphs showing the distribution of gender among Hindi and English Users. It clearly shows skewness in gender for English users whereas in the case of Hindi users it shows an approximate normalization among the genders.

A total of 436 people took part in the English language based survey, and 126 took part in the Hindi language based survey. This plot was created based on the survey conducted on the WashKaro app. The analysis of this plot suggests that the number of English users are more than the number of Hindi users. It also depicts that the overall number of male users is more than female users. A key insight observed from the data depicted that Hindi speaking female users (33% of total Hindi speaking users) were more than English speaking female users (11% of total English speaking users). The census of India 2011, highlighted the disparity of literacy rates across genders with 82.14% literacy rates amongst Indian males and 65.46% literacy rates in Indian females [13]. This underscores the fact that using local languages empowers the sections of the population that might not have otherwise access to the information. Also, as highlighted by previous work, teaching interventions to women is an effective method of mitigating diseases, given their greatest willingness to devote time towards health and survival [14].

**Target population: Users who reported higher than expected incidence**

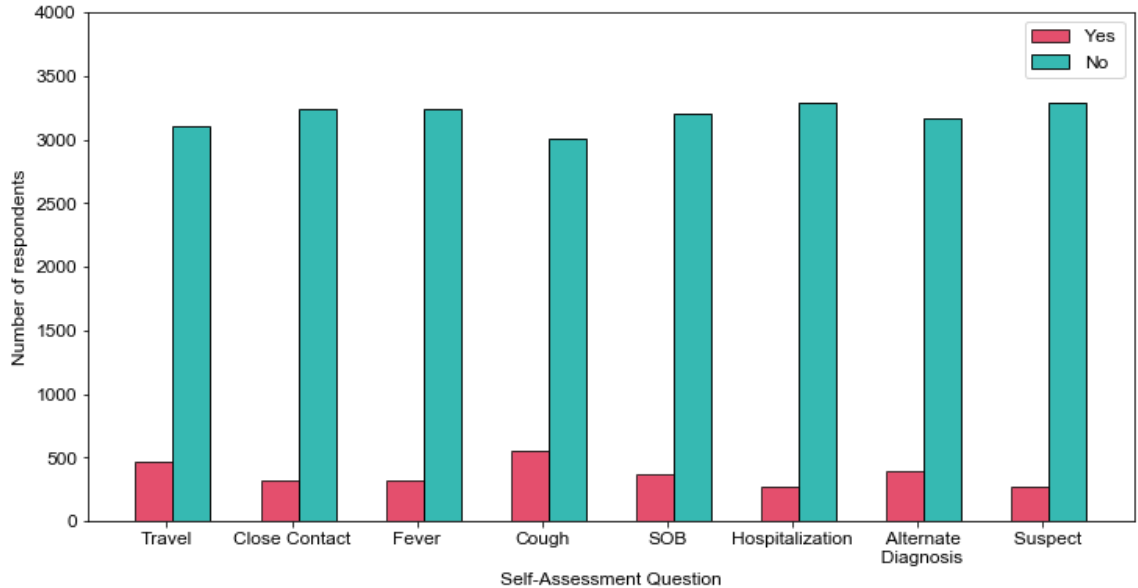

Figure 7. Analysis of Self Assessment. A simple user-level self-assessment has been deployed in order to enable the general population to perform self-assessment and to identify the population at risk which can be used as an effective screening. This vividly states that the person tested for self-analysis was higher than expected numbers.

Based on the data collected from the symptom self-assessment we can distil the general population so that we can trace the potential number of positive patients who are at risk of COVID-19 and can also help in identifying patients who are already suffering from this disease. The question has been framed in such a way that potential threats are identified at the early stage so that we can mitigate the chances of community transmission. Out of 3567 respondents, 276 (7% of respondents) were found to be suspect cases according to their responses, while 3291 were non-suspect. 467 (13.09 %) of the respondents reported a travel history to locations reporting community transmission of COVID-19, 326 (9.13%) reported close contact with COVID-19 positive patient, 323 (9.05% of respondents) reported fever, 556(15.58%) reported cough, 367(10.28%) reported shortness of breath (SOB), 277 (7.77%) reported that they required Hospitalization and 395 (11.07%) respondents reported there was an alternate diagnosis for their condition.

## Discussion

Over the years mHealth and machine learning have made significant contributions in the medical domain. In the case of COVID-19, where reliable therapeutic strategies are still under experimentation, the role of such mHealth and machine learning interventions is critical. Timely dissemination of trusted byte-sized information is extremely instrumental in mitigating the infodemic. With the majority of the world population staying at home, the increased amount of digital consumption opens up the scope to deploy these techniques as an effective social intervention to mitigate the infodemic by delivering the right information to the right people at the right time. When organizations across the globe are proactively testing various strategies to address the issue at hand, open-source software will play a vital role in such scenarios at a global stage to mitigate pandemics and infodemics at the root level. We provide an open-source template featured with functions like Symptom Self-Assessment, Notification amplifier which notifies the user for washing hands, which are required for fighting against epidemics and pandemics. This also helps in the propagation of the right information hence proper management of infodemic can be done without misleading the masses in crucial scenarios. The enhancement of a few features within the WashKaro application can help in serving as an effective intervention for the government and policymakers. Detailed questions can be formulated targeting the at-risk users identified using symptom self- assessment, which can be incorporated into the existing framework followed by the higher authorities and medical workers for predicting the suspects of COVID-19 at an early stage. After identification and testing of the at-risk population, our analysis can be extended to predict patients who have chances of being at risk in the near future from the definitive set of questions, based on the priority of each question. We can present this data to the suitable administration and decision-makers for taking effective measures against such individuals at an early stage. For the Suspect Cases, we can administer a second questionnaire to further stratify the risk of acute respiratory distress syndrome (ARDS) and septic shock by assessing the severity of symptoms and looking for identified risk factors like age and pre-existing comorbidities that are not included in the WHO Interim Guidance. This can aid in making decisions regarding home quarantine against hospital admission. The app can also be used to identify other suspect cases in the same household. Further, to assist the government authorities to identify those requiring testing for COVID-19, we can ask for contact details of the Suspect Cases with informed consent and relay them to the appropriate government authorities to enable targeted testing. A follow up of the suspect cases through push notifications, advising testing and recording test results, can help ensure that complacence does not set in.

Finally, there are some limitations to the study that have been conducted prior to the revamp of the application. All applications with COVID-19 information were removed in association with the guidelines regarding COVID-19 related applications on the Google Play store. The time frame of the case study for COVID-19 was shortened due to this reason. Now, WashKaro is a generic public health intervention suite that is currently structured to address Tuberculosis(TB) as it is the leading cause from a single infectious agent [15]. The Application currently features multilingual success-stories, FAQs, Mythbusters, a TB based chatbot, a TB awareness quiz, a Twitter Analysis screen for TB related tweets, and insights from tweets related to Tuberculosis. Secondly, due to data privacy policies, we could not obtain granular information on individuals' locations, such granularity of data could have helped articulate user studies in a more diverse way. The development of innovative approaches while protecting individual data yet gathering useful inference is an active area of research, and our further work will address this limitation in various public health scenarios. Therefore, we conclude that the role of digital health interventions in the form of systems articulating vetted messages needs to be explored effectively dealing with public health challenges, both during health emergencies and normal times addressing the Sustainable Development Goals (SDGs) put forward by WHO.

## Acknowledgements

This work was partly supported by the Wellcome Trust/DBT India Alliance Fellowship IA/CPHE/14/1/501504 awarded to Tavpritesh Sethi. Tavpritesh Sethi also acknowledges support from the Center for Artificial Intelligence at IIIT-Delhi. We acknowledge Bhavika Rana for her inputs on App design and Prof. Rakesh Lodha, AIIMS New Delhi, for his clinical inputs.

## Conflicts of Interest

No conflicts of interest declared.

## Abbreviations

WHO: World Health Organization
WASH: Water Sanitation Hygiene
EPI-WIN: WHO's Information Network for Epidemics
AI: Artificial Intelligence
NLP: Natural Language Processing